\documentclass[10pt, conference]{IEEEtran}
\IEEEoverridecommandlockouts
\usepackage{cite}
\usepackage{amsmath,amssymb,amsfonts}
\usepackage{algorithmic}
\usepackage{graphicx}
\usepackage{textcomp}
\usepackage{xcolor}

\usepackage{package}

\begin{document}



\title{On the Impact of 3D Visualization of Repository Metrics in Software Engineering Education}

\author{
\IEEEauthorblockN{Dario Di Dario}
\IEEEauthorblockA{\textit{Computer Science Department} \\
\textit{University of Salerno}\\
Fisciano, Italy}
\and
\IEEEauthorblockN{Stefano Lambiase}
\IEEEauthorblockA{\textit{Computer Science Department} \\
\textit{University of Salerno}\\
Fisciano, Italy}
\and
\IEEEauthorblockN{Fabio Palomba}
\IEEEauthorblockA{\textit{Computer Science Department} \\
\textit{University of Salerno}\\
Fisciano, Italy}
\and
\IEEEauthorblockN{Carmine Gravino}
\IEEEauthorblockA{\textit{Computer Science Department} \\
\textit{University of Salerno}\\
Fisciano, Italy}
}

\maketitle

\begin{abstract}
\underline{Context:} Software development is a complex socio-technical process requiring a deep understanding of various aspects. In order to support practitioners in understanding such a complex activity, repository process metrics, like number of pull requests and issues, emerged as crucial for evaluating CI/CD workflows and guiding informed decision-making. The research community proposed different ways to visualize these metrics to increase their impact on developers' process comprehension: VR is a promising one. Nevertheless, despite such promising results, the role of VR, especially in educational settings, has received limited research attention.
\underline{Objective:} This study aims to address this gap by exploring how VR-based repository metrics visualization can support the teaching of process comprehension.
\underline{Method:} The registered report proposes the execution of a controlled experiment where VR and non-VR approaches will be compared, with the final aim to assess whether repository metrics in VR's impact on learning experience and software process comprehension. By immersing students in an intuitive environment, this research hypothesizes that VR can foster essential analytical skills, thus preparing software engineering students more effectively for industry requirements and equipping them to navigate complex software development tasks with enhanced comprehension and critical thinking abilities.
\end{abstract}



\section{Introduction}

Developing software is a complex socio-technical activity, where the software lifecycle encompasses multiple stages that need to be correctly managed in order to ensure the project success~\cite{baum2017choice}. Mitigating such complexity necessitates a thorough understanding of the software process itself. Thus, the ability to \textbf{comprehend and analyze the development process} is critical not only in industry and open-source practices~\cite{caserta2010visualization, bacchelli2013expectations, macleod2017code, visualization} but also in software engineering (SE) education~\cite{patani, hsing2019using}. To facilitate a deeper understanding of the development process, research communities have introduced various tools, with process metrics emerging as key instruments. Software process metrics provide quantitative measures for assessing, monitoring, and enhancing the effectiveness and efficiency of software development workflows. Particularly, in Continuous Integration and Continuous Delivery (CI/CD) environments, \textbf{repository (process) metrics}, such as the number of pull requests, issues, and commits offer insights into the health of software communities that are essential to enable contributors to make informed decisions that enhance process quality and contribute to improving overall well-being within the development environment~\cite{gousios2008measuring}.

To support practitioners into understanding those metrics, researchers started to investigate \textbf{visualization strategies}~\cite{moreno2024influence, langelier2005visualization, wettel2007visualizing, wettel2008visually}. Leveraging principles from perception, cognition, and graphical design, these strategies encode data values and attributes through visual elements and properties, which has been shown to improve comprehension of complex information, facilitate faster communication compared to textual or numerical data, and reduce cognitive load for easier analysis. The literature thus indicates that these strategies \textbf{support practitioners in comprehending software processes}, a crucial skill for key activities in both industry—such as software maintenance~\cite{caserta2010visualization} and code review~\cite{bacchelli2013expectations, macleod2017code, visualization}—and education, particularly in teaching SE~\cite{patani, hsing2019using}.

Metric visualization is typically performed using a keyboard and mouse in a 2D environment (e.g., a monitor). Nevertheless, their effectiveness in improving the program comprehension process has led to the development of innovative tools that incorporate advanced technologies like \textbf{virtual reality (VR)}~\cite{moreno2023codecity, moreno2022babiaxr, moreno2024software}. Although still relatively new, VR-based tools for visualizing repository metrics are already proving valuable for enhancing development activities. Despite increasing research interest in streamlining processes for IT practitioners~\cite{moreno2023codecity, moreno2022babiaxr, moreno2024software} and academic researchers~\cite{garousi}, relatively little focus has been placed on \textbf{educational applications}. Specifically, while the importance of process comprehension activities, such as those supported by CI/CD platforms, is well-established for post-university careers in SE~\cite{patani, hsing2019using}, there is \textbf{limited research that explores how VR-based software visualization techniques could support teaching these essential skills for software engineers.}

Addressing this gap is essential for several reasons. First, the increasing pervasiveness of open-source in the software development environment has led to widespread adoption of CI/CD platforms—and consequently, repositories—as standard practice. Since repository metrics are inherently part of these systems, equipping students with the skills to understand these metrics is crucial for the future of the field. Second, comprehension of process metrics is increasingly valuable as large organizations, including Microsoft and Google, integrate practices rooted in these skills, such as code review~\cite{chong2021assessing, hundhausen2009integrating, wang}. Third, teaching software metrics comprehension has been shown to benefit students by fostering critical analytical, discipline-specific, and collaborative skills~\cite{bisant1989two, davila2021systematic, patani, hsing2019using}—core competencies within the socio-technical landscape of software development~\cite{storey2020software, hoda2021socio}. Additionally, research in SE has shown that incorporating VR into process understanding can shorten learning curves and enhance creativity, providing substantial benefits for university-level education~\cite{elliott2015virtual}.


Given (1) the promising results achieved through integrating VR for software process comprehension~\cite{moreno2023codecity, moreno2024influence}, (2) the documented importance of teaching software comprehension to future professionals~\cite{patani, hsing2019using}, and (3) the need to enhance the teachability of this critical skill, this confirmatory registered report builds on the work of Moreno-Lumbreras et al.~\cite{moreno2024software} by \textbf{examining whether VR, when used to visualize repository metrics, can improve software engineering students' educational experiences in comprehending development processes}. VR may indeed provide students with a more intuitive grasp of complex concepts, thereby enriching the learning experience and better equipping them for real-world software engineering challenges.

\section{Background and Related Studies}

This section provides a brief overview of the foundational concepts necessary to understand this research.

\subsection{Software Visualization}

Software visualization plays a vital role in comprehending programs, particularly during maintenance and evolution activities, as it helps interpret complex data and supports informed decision-making. Various techniques exist, each adopting a different approach to visualization. For instance, \textit{graph-based} methods depict structural relationships and system changes using nodes and links, creating a network of interconnected elements. In contrast, \textit{notation-based} approaches, such as UML diagrams, utilize formal models to represent system modifications in a structured way. \textit{Matrix-based} techniques focus on tracking the version history and relationships between different modules, providing insights into the evolution of specific components over time. Meanwhile, \textit{metaphor-based} methods offer a more intuitive approach by employing visual metaphors, like visualizing the software as a 3D city, to make system evolution more engaging and easier to understand~\cite{software_visualization}.

These techniques serve different purposes, depending on the level of abstraction and the aims. Some approaches are \textit{artifact-centered}, aiming to show how individual elements or files change over time. Others are \textit{metric-centered}, designed to highlight variations in quality parameters across versions. There are also \textit{feature-centered} techniques that assess the impact of modifications on particular system features, and \textit{architecture-centered} approaches that visualize the evolution of the software's overall structure, including component dependencies and interactions \cite{software_visualization}.

Software repositories often provide the primary data source for these visualizations, documenting the project's entire development history, from minor edits to major releases. In some cases, the source code itself is used, either alongside repository data or independently, to offer a more detailed view of the software's progression and transformations \cite{software_visualization}.

\subsection{Software Visualization and Virtual Reality}

A significant area where software visualization proves beneficial is in \textit{virtual reality (VR)}. In the realm of VR, a virtual environment is a digitally created space that enables users to interact with simulated settings, whether realistic or imaginary, offering an immersive experience by simulating a sense of physical presence in both real-world and fantastical scenarios \cite{hale2014handbook, blom2014design}. Virtual reality visualization has both strengths and weaknesses, as pointed out by Teyseyre et al. \cite{teyseyre}. On the positive side, it allows a larger amount of data to be represented compared to 2D methods. This can be particularly useful in fields where analyzing a large volume of data quickly is required. For example, in the context of software engineering, the third dimension is used to show the complex relationships and multiple attributes of a software project. On the negative side, many users may find it difficult to use this technology, as they are more familiar with 2D tools and are not accustomed to wearing VR headsets, motion sensors, or haptic devices \cite{hale2014handbook, blom2014design}. Additionally, the development of VR applications demands advanced technical skills to create user-friendly 3D graphics and interfaces. It also requires more computational resources than 2D applications.

Among the various software visualization techniques mentioned, the \textit{software city metaphor} is one of the most widely used within virtual reality environments. By visualizing software as a city, where buildings and streets represent various components and relationships, users gain a more engaging and intuitive understanding of system evolution. This approach has proven successful in making complex software data more accessible and interactive, enhancing the overall experience of software maintenance and exploration \cite{merino, merino2, fittakau, rudel}. 

More in detail, the concept of software visualization with city metaphor was first introduced by Wettel et al. \cite{wettel2007visualizing}, leveraging the idea that software elements can be statically depicted as various city components, such as buildings and roads, simplifying overall software comprehension. Over time, the same idea has evolved and adapted to different contexts by Merino et al. \cite{merino}, including virtual reality and augmented reality \cite{merino2}.
Building on these works, various adaptions have enriched the concept of city metaphor and software visualization. Kobayashi et al. \cite{kobayashi} represented software architectures with buildings as packages and lines as dependencies. Yano et al. \cite{Yano} extended this concept by adding metrics to capture software characteristics based on dependencies. Similarly, Misiak et al. \cite{misiak} with \textit{IslandViz} visualized architectures as islands, depicting dependencies as connections and arrows to indicate hierarchy. These adaptations offer diverse ways to illustrate software complexity and relationships.
Finally, Moreno-Lumbreras et al. \cite{moreno2024software}, building on top of previous work, provide a more in-depth, dynamic 3D-visualization in the context of software repository metrics, that are able to characterize development process, proving the ability to comprehend and analyze interconnected metrics. Indeed, the study focuses on understanding whether VR can help users analyze processes like code review and issue tracking more effectively. To do this, the authors conducted an experiment where participants used both a 2D dashboard setup based on Kibana\footnote{Kibana: \url{https://www.elastic.co/kibana}}, and an immersive VR environment created with BabiaXR\footnote{BabiaXR: \url{https://babiaxr.gitlab.io}} to explore software data. The study included 32 participants from academic and industrial backgrounds, all of whom completed tasks in both environments. The research assessed the accuracy of participants' responses and their efficiency in completing tasks. The findings indicated that accuracy levels were comparable in both settings. However, virtual reality seemed to provide an edge in handling more intricate tasks that necessitated rapid examination of interconnected visualizations. 
All the work mentioned above poses a significant basis for contribution in the educational field because academic courses should provide the necessary tools and methods to prepare students for real-world scenarios. By integrating advanced visualization techniques, such as those proposed by Moreno-Lumbreras et al., educators can bridge the gap between theoretical knowledge and practical skills, enabling students to better understand, analyze, and monitor complex software systems.

\subsection{Software Visualization for Software Engineering Education}
Understanding software development processes is vital in software engineering education. Foundational studies by Patani et al. \cite{patani} and Hsing et al. \cite{hsing2019using} highlight the importance of teaching process metrics like pull request and commit, especially in CI/CD environments, fostering critical thinking, collaboration, and industry readiness. Similarly, Bisant and Lyle \cite{bisant1989two} and Dávila and Rodríguez \cite{davila2021systematic} emphasize that analyzing process metrics strengthens teamwork and socio-technical competencies.
Elliott et al. \cite{elliott2015virtual} show that VR accelerates learning and fosters creativity, offering an immersive way to teach complex topics like repository metrics. These findings support the integration of innovative tools to improve learning outcomes.

\steSummaryBox{\faList \hspace{0.05cm} Related Work: Summary and Research Gap.}{Current literature predominantly focuses on testing VR-based software visualization thinking to the industry sector~\cite{moreno2024software, moreno2022babiaxr}, leaving a notable gap in its exploration within educational contexts. Given that repository metrics visualization—through tools like CI/CD—has been effective in SE courses~\cite{patani, hsing2019using} and is fundamental for conveying core software development concepts~\cite{bisant1989two, davila2021systematic, patani, hsing2019using}, an investigation into VR's application for educational process comprehension appears worthwhile. 
}

\section{Goal of the Study}

The primary \textit{goal} of this study is to assess whether visualizing repository metrics within a VR environment enhances students' learning experience and comprehension of software processes, in comparison to traditional 2D methods. We decided to explore three dimensions of learning experience, i.e., (1) learning outcome, (2) students performance in classroom tasks, and (3) quality of the experience. The \textit{purpose} is to explore novel strategies for enhancing teaching activities related to code review, software maintenance, and, more broadly, software comprehension. This investigation focuses on the \textit{perspective} of university students who are seeking to deepen their understanding of software-related activities, aiming to enter the professional world better prepared.

The \textbf{leading hypothesis} of this study is the following: \textit{Since 3D virtual environments allow for more immersive and comprehensive visualization of repository metrics, their use could support software engineering students in performing and understanding process comprehension, ultimately enhancing their learning experience.}

The hypothesis is supported by recent literature in the software engineering (SE) field, which demonstrated that VR could benefit professionals in SE, support maintenance and code review activities~\cite{moreno2022babiaxr, moreno2024influence}, improve software project comprehension~\cite{moreno2023codecity}, and stimulate creativity and learning~\cite{elliott2015virtual}. Nevertheless, it is disputable because VR still suffers—since it is still an emerging technology—on various problems, particularly from the comfort and realism side.

\section{Hypotheses and Research Questions}

Aiming to test the crafted hypothesis, we identify a set of dimensions and formulate research questions (RQs) that are introduced and motivated in the following.

The first dimension of the student's learning experience that we plan to investigate is the \textbf{learning outcome}, that can be defined as ``\textit{the learning results that a student is expected to know, understand, and demonstrate at the ending of learning experience}''\cite{adam_learning_outcome}.
For this reason, we formulate the following research question:
\steResearchQuestionBox{\faQuestionCircleO \hspace{0.05cm} \textbf{RQ\textsubscript{1}}: \textit{Is the \underline{learning outcome} gained through repository metrics visualization by students in a VR environment comparable to that gained in a traditional on-screen setting?}}

We formalize the RQ\textsubscript{1} with the following hypotheses:
\begin{itemize}[leftmargin=0.35cm] 
    \item \textbf{H1\textsubscript{0}}: The use of repository metrics visualization provides no significant differences in learning outcomes for software process comprehension between students using VR and those using a traditional on-screen environment.
    \item \textbf{H1\textsubscript{a}}: The use of repository metrics visualization provides significant differences in learning outcome for software process comprehension between students using VR and those using a traditional on-screen environment.
\end{itemize}

The second dimension that we plan to explore is the students' \textbf{performance in classroom tasks}. Drawing on the foundational work of Moreno-Lumbreras et al.\cite{moreno2024software}, this study aims to assess students’ performance in terms of accuracy and efficiency while conducting maintenance tasks outlined by Sillito et al.\cite{sillito2006questions}. While the original work targeted practitioners, this research aims to expand and complement it by examining similar metrics—\textit{accuracy} for correctness and \textit{efficiency} for time to completion—in an academic setting. Accordingly, the following two research questions were formulated:


\steResearchQuestionBox{\faQuestionCircleO \hspace{0.05cm} \textbf{RQ\textsubscript{2}}: Are the answers provided by students, using the repository metrics visualization in a VR environment, comparable in \underline{correctness} to those provided in a traditional on-screen environment?}

\steResearchQuestionBox{\faQuestionCircleO \hspace{0.05cm} \textbf{RQ\textsubscript{3}}: \textit{Are the answers provided by students, using the repository metrics visualization} in a VR environment, comparable in \underline{time to completion} to those provided in a traditional on-screen environment?}
We formalize the two RQs in the following hypotheses:
\begin{itemize}[leftmargin=0.35cm]
    \item The use of repository metrics visualization provides no significant differences in the (\textbf{H2\textsubscript{0}}) correctness and (\textbf{H3\textsubscript{0}}) time to completion of answers provided by students in a VR environment compared to those provided in a traditional on-screen environment.

    \item The use of repository metrics visualization provides significant differences in the (\textbf{H2\textsubscript{a}}) correctness and (\textbf{H3\textsubscript{a}}) time to completion of answers provided by students in a VR environment compared to those provided in a traditional on-screen environment.
\end{itemize}

Last, using VR in an educational context requires major attention to a plethora of aspects compared to the one in a practical environment~\cite{pirker2020virtual, hamilton2021immersive, fernandes2022evaluating, akbulut2018effectiveness}. Indeed, the \textbf{students' quality experience}, measured through factors validated in the literature like immersiveness and perceived workload are crucial during the learning activity; moreover, we argue that when using innovative technologies—like VR in education—a more comprehensive view of the experience is needed in order to provide valuable results. Thus, we formulated the following last research question:

\steResearchQuestionBox{\faQuestionCircleO \hspace{0.05cm} \textbf{RQ\textsubscript{4}}: \textit{Is the \underline{quality of students' learning experience} when visualizing repository metrics in a VR environment comparable to that in a traditional on-screen setting?}}
We formalize RQ\textsubscript{4} in the following hypotheses:
\begin{itemize}[leftmargin=0.35cm]
    \item \textbf{H4\textsubscript{0}}: The use of repository metrics visualization provides no significant differences in the factors related to the students' quality experience in a VR environment compared to those in a traditional on-screen environment.
    \item \textbf{H4\textsubscript{a}}: The use of repository metrics visualization provides significant differences in the factors related to the students' quality experience in a VR environment compared to those in a traditional on-screen environment.
\end{itemize}


\section{Design}
To answer the RQ(s) formulated in previous section, we will conduct a controlled experiment, following the \textit{ACM SIGSOFT Empirical Standards} and Wohlin et al. guidelines~\cite{wohlin2012experimentation}.


\subsection{Experimental Setup}
To investigate whether virtual reality can improve process comprehension in education, the experiment will take place in two distinct environment settings:

\begin{description}[leftmargin=0.3cm]
    \item[2D On-Screen:] Participants will use a computer with a web browser displaying a \textsc{Kibana} dashboard that includes various panels for visualizing repository metrics.
    \item[Virtual Reality:] Participants will engage with a VR environment through a Meta Quest Pro headset, using the VR browser to access a \textsc{BabiaXR} presenting repository metrics on interactive 3D panels, allowing them to explore and interact with data immersively.
\end{description}

In each setting, participants will perform identical tasks across both modalities and on two different projects. Tasks will be recorded via screencasting, and participants' verbal responses will be transcribed for analysis. A supervisor will be present to oversee the process, assist as necessary, and communicate with participants as needed.

Initially, in the study conducted by Moreno-Lumbreras \cite{moreno2021rr} (that inspired this study), a ``One Factor, Two Level'' design was used, where each participant completed only a subset of the specified tasks. However, this design was later modified to a ``within-subject'' format, allowing all participants to perform tasks across both settings of the independent variable. This transition, highlights the importance of accounting for learning effects, which can occur when participants repeat similar tasks across experimental conditions. 
To ensure the validity of the assessment of students’ learning outcomes—defined as the knowledge, skills, or competencies expected to be gained from the learning experience—an AB/BA crossover design was adopted. This approach, as outlined by Vegas et al. \cite{vegas2015crossover}, was considered particularly well-suited to the study because potential learning effects, such as unintended performance improvements resulting from repeated exposure to similar tasks, are mitigated. The order in which groups engage with the 2D and 3D environment conditions was randomly alternated to minimize biases that could compromise the results. Additionally, a rest period was included between the two sessions to reduce carryover effects and fatigue. Through this careful consideration of experimental design, the findings are ensured to accurately reflect the effectiveness of each environment in supporting learning outcomes.
This design enhances comparison reliability by allowing each participant to act as their own control across different conditions.

\subsection{Variables}
To conduct the experiment, the variables to be measured are outlined below. Nevertheless, the independent variable—i.e., the environment setting—is common between all the RQs. 

\begin{description}
    \item[Independent Variable:] ``Environment Settings''
    \begin{itemize}
        \item \textbf{Description:} Environments where participants are requested to perform specific software engineering tasks.
        \item \textbf{Scale:} 2D On-Screen AND VR
    \end{itemize}
\end{description}

Regarding the first research question—focusing on the \textit{learning outcome}—we take inspiration from works exploring using instruments for repository metrics (e.g., CI/CD platforms) in an educational environment~\cite{hsing2019using, patani}. We then decided to measure learning outcome by mean of a (1) self-reported experience scales (implemented using a questionnaire and inspired by the literature~\cite{hsing2019using, patani}) and (2) evaluation quiz on software projects comprehension course.

\begin{description}[leftmargin=0.3cm]
    \item[Dependent Variable:] “Self-assessed learning on process comprehension”
    \begin{itemize}
        \item \textbf{Description:} How much the student feels they have learned regarding concepts of process comprehension.
        \item \textbf{Scale:} Questionnaire adapted by the literature~\cite{hsing2019using, patani} and using Likert Scale of Agreement.
        \item \textbf{Operationalization:} Average of the answers.
    \end{itemize}
    
    \item[Dependent Variable:] ``Self-assessed perceived benefits''
    \begin{itemize}
        \item \textbf{Description:} How prepared the student feels for the job market and for performing software comprehension tasks in a real-world context.
        \item \textbf{Scale:} Questionnaire adapted by the literature~\cite{hsing2019using, patani} and using Likert Scale of Agreement.
        \item \textbf{Operationalization:} Average of the answers.
    \end{itemize}
    
    \item[Dependent Variable:] ``Comprehension''
    \begin{itemize}
        \item \textbf{Description:} What knowledge the student actually possesses about software comprehension activities.
        \item \textbf{Scale:} Quiz questionnaire (with both closed and open-ended questions) aimed at measuring learning.
        \item \textbf{Operationalization:} Closed-ended answers are scored as either correct (1), incorrect (-0.25), or not provided (0). Open-ended answers are scored in the range [0,1].
    \end{itemize}
\end{description}
The operationalization of \textit{Comprehension} is based on our university’s exam framework. This framework typically uses five multiple-choice options for each question, where incorrect answers receive a negative score. As demonstrated by Espinosa et al. \cite{espinosa}, such approach ensures that the probability of selecting a correct answer by guessing is zero.

Inspired by the contribution provided by Moreno-Lumbreras et al. \cite{moreno2024software}, we define the dependent variables related to the second and third research questions. Specifically, we will assess the \textit{accuracy} and \textit{efficiency} of students performing the SE tasks as suggested by Sillito et al. \cite{sillito2006questions} in the two environments. On the one hand, accuracy can be defined as the deviation between a correct (scored as 1) and an incorrect answer (scored as 0) for specific tasks, with scores within this range indicating how partially correct the answer is. On the other hand, efficiency in completing a specific task can be defined as the time taken to perform the task, measured from the moment participants indicate they are ready to begin until a final response is provided. This measure is recorded independently of whether the response is correct or incorrect.

\begin{description}[leftmargin=0.3cm]
    \item[Dependent Variable] ``Accuracy''
    \begin{itemize}
        \item \textbf{Description:} How closely the student response aligns with the correct answer.
        \item \textbf{Scale:} Float
        \item \textbf{Operationalization:} Answers scored in the range [0,1].
    \end{itemize}
    
    \item[Dependent Variable:] ``Efficiency''
    \begin{itemize}
        \item \textbf{Description:} Time employed by participants to complete each task assigned.
        \item \textbf{Scale:} Real Number
        \item \textbf{Operationalization:} Number of seconds to understand the question and provide the answer. 
    \end{itemize} 
\end{description}

In our fourth research question, we plan to assess the overall experience of students interacting in a VR environment compared to a non-VR setting. Drawing on literature regarding VR in educational contexts~\cite{pirker2020virtual, hamilton2021immersive, fernandes2022evaluating, akbulut2018effectiveness}, we chose to evaluate the quality of experience through the following validated dimensions: (1) \textit{engagement}, (2) \textit{satisfaction}, (3) \textit{perceived workload}, (4) \textit{immersiveness}, and (5) \textit{presence}. The engagement dimension will be assessed using five factors \cite{wang_engagement}, namely: \textit{affective engagement}, \textit{behavioral engagement–compliance}, \textit{behavioral engagement–effortful}, \textit{cognitive engagement}, and \textit{disengagement}. Satisfaction will be measured using the \textit{Net Promoter Score} \cite{kaplan_nps} and validated as an academic indicator by Baehre et al. \cite{baehre_nps}. Perceived workload will be evaluated using the \textit{NASA-TLX} to calculate the task load index \cite{hart_nasa}. Immersiveness and presence will be measured using the metrics proposed by Witmer et al. \cite{witmer_presence}, adapted to the study’s context. These tools capture the degree of environmental engagement, sense of presence, and user focus. Each of these will be measured using validated and adapted questionnaires, employing Likert scales of agreement ranging from 1 to 5, 1 to 7, and 1 to 10.

Moreover, we identify some \textbf{confounding variables} that could influence the outcome of our experiments. First, \textit{experience with \textsc{Kibana}} will be assessed by asking participants to self-report their familiarity with the dashboard, categorizing their experience into levels of “None,” “Small,” “Medium,” and “High.” Participants indicate their approximate hours spent using the tool and the context in which they applied it. Similarly, \textit{experience with Virtual Reality} will be quantified by asking participants to describe their familiarity with VR technology on a categorical scale from “None” to “High,” including indicative usage hours and specific contexts. Additionally, \textit{programming experience} will be captured as an integer representing the number of years participants had engaged in academic programming activities, providing a numeric basis for comparison. It is also important to evaluate, as done by Moreno-Lumbreras et al. \cite{moreno2024software}, whether participants have any prior experience with the projects used in the experimental tasks. Finally, \textit{experience in data visualization} will be evaluated by inquiring about participants’ familiarity with data visualization tools and techniques, with responses recorded as integer values.

\subsection{Participants}
A key difference between their study and ours lies in the participants' backgrounds, knowledge, and expertise. Moreno-Lumbreras et al. recruited participants with diverse professional backgrounds, including developers, system architects, telecommunications engineers, consultants, and master’s students with knowledge related to software development. In contrast, this study will focus on bachelor’s and master’s students from our university, with academic experience in software development and software engineering. Building on this difference, a preliminary sample size estimation was conducted using \textit{G*Power} to provide an initial assessment of the required number of participants. The results indicated a minimum sample size of approximately 130 individuals. However, it remains uncertain whether this number can be achieved. For this reason, the experiment will involve at least 32 participants, aligning with the sample size used in the controlled experiment conducted by Moreno-Lumbreras et al.~\cite{moreno2024software}. It is worth noting that independently from the number of participants, students were selected exclusively from our university for two main reasons. First, as highlighted by Baltes and Ralph \cite{baltes2022sampling}, identifying a group of participants that fully represents the entire population is highly challenging. Second, including students from other universities would require significant effort and collaboration, without ensuring representation of the broader student population referenced in \cite{baltes2022sampling}.

The participants’ software development knowledge is essential to ensure familiarity with foundational principles and logic in development. Equally important are SE skills, which not only confirm participants’ development expertise but also provide a rigorous sample of individuals experienced with repository metrics and tools like \textsc{Git} and \textsc{GitHub}. Moreover, experience in SE projects could provide a sample of participants with limited experience in teamwork and communication toward a common goal. Although the SE course offers participants a basic understanding of the code review process, they may lack practical experience in applying these skills. 

To control potential confounding variables, participants will complete a demographic form. This form will assess their background and familiarity with VR and \textsc{Kibana}. However, following the approach of Moreno-Lumbreras et al. \cite{moreno2024software}, participants will not be grouped based on their prior experience to avoid introducing any bias.

\subsection{Projects, Tools, and Tasks}
The experiment will use two open-source projects, \textit{CHAOSS}\footnote{CHAOSS: \url{https://chaoss.community}} and \textit{OpenShift}\footnote{OpenShift: \url{https://docs.openshift.com/}}, which are managed by separate communities and contain extensive datasets related to open-source repository metrics. These projects provide rich sources of data for analyzing software development process. CHAOSS, a project under the Linux Foundation, focuses on creating metrics and tools to assess the health of open-source communities. It offers tools like GrimoireLab\footnote{GrimoireLab: \url{https://chaoss.github.io/grimoirelab/}}, which visualizes repository data on thousands of issues, pull requests, and contributors across multiple organizations. OpenShift, managed by Red Hat, focusing in its extensive community, involving thousands of contributors and organizations, supporting in collecting a wide range of repositories, issues, and pull requests.

For data visualization, we will use two tools: \textsc{Kibana} and \textsc{BabiaXR} \cite{moreno2022babiaxr}. Kibana will display repository metrics on a 2D screen with interactive dashboards, allowing participants to examine data in a traditional format. In contrast, the VR environment, BabiaXR, will present the same metrics in a 3D virtual space, potentially enhancing their understanding. 
Participants will complete the same five maintenance tasks defined by More-Lumbreras et al. \cite{moreno2024software} in both VR and on-screen environments. Based on the framework by Sillito et al. \cite{sillito2006questions}, tasks will evaluate the specified dependent variables across repository metrics in both visualization methods.

The tasks will focus on repositories data from CHAOSS and OpenShift projects. Data will be visualized at multiple levels within both \textsc{Kibana} and \textsc{BabiaXR}: (1) project level, (2) organization level, (3) contributor level, (4) repository level, and (5) subproject level.
For example, an organizational performance task might require participants to identify open pull requests or issues across projects and measure the time required to close them. A project performance task could involve finding key contributors and determining which members have contributed most to specific pull requests or issues. This setup aims to compare participants’ performance and insights when using different data visualization formats.

\subsection{Execution Plan}
Since our experiment will be a within-subject with a crossover design, the execution will be arranged on 4 different days. Specifically, the students' sample will be randomized into 4 groups without balancing their skills provided with the demographic questionnaire. Each group will be executed on a different day. Below is the workflow that will be repeated for each experimental group.
\begin{description}[leftmargin=0.3cm]
    \item[Welcome Part.] Sessions start with a welcome to create a comfortable environment for participants. The researcher briefly outlines the experiment's goals, explains the tasks, and introduces the two methods that will be used. Students are reminded of data confidentiality and encouraged to engage thoughtfully. Finally, participants will take part in a VR training session to calibrate the headset and become familiar with the virtual environment. 
    \item[The Software Visualization Tasks.] In the main session, students complete data visualization tasks, starting in either VR or 2D on-screen scenes, depending on their assigned group. Performance metrics such as accuracy and efficiency will be recorded to answer RQ\textsubscript{2} and RQ\textsubscript{3}. After finishing the first set, students switch to the other modality and perform the same tasks on a different project.
    \item[Learning Outcome and Experience.] After completing the tasks, students will complete the learning outcome questionnaire to address RQ\textsubscript{1} and the overall students' experience questionnaire to address RQ\textsubscript{4}. 
\end{description}

\subsection{Analysis Plan}
In accordance with Moreno-Lumbreras et al. \cite{moreno2024software}, we will employ a \textit{Mixed Linear Model} as the primary statistical technique to address the RQs. This modeling approach is particularly advantageous given our crossover design, as it facilitates the inclusion of control variables such as treatment order and individual participant differences.
However, several linear models will be developed for each research question, each tailored to a specific dependent variable. This approach ensure that the complexity of each linear model remains comparable to that employed by Moreno-Lumbreras \cite{moreno2024software}. Moreover, Mixed Linear Model also require numerous checks (including test for multicollinearity), adding further complexity. To avoid providing inaccurate results by forcing statistical model that might not be robust with certain data, we plan to rely on traditional statistical methods, such as \textit{T-test}, and \textit{ANOVA}, grouping the samples based on the experimental settings.
However, before applying the statistical test, the \textit{Shapiro-Wilk} test will be used to assess whether the null hypothesis of normality can be rejected. In addition, a visual inspection of the data will be done to support the results. In instances where the residuals fail to conform to a normal distribution and transformations of the data do not rectify the issue, we will pivot to the utilization of a \textit{Generalized Linear Mixed Model} (GLMM). Finally, an additional step could be applied for the RQ$_1$ and RQ$_4$: as specified in the previous section, the learning outcome and the overall experience of the students will be calculated by means of different values. For this reason, before running the mixed linear model, we will plan to perform the \textit{Exploratory Factor Analysis} (EFA) in order to assess constructs validity.

\subsection{Ethical Considerations}

Since our experiment involves human participants, we take into account ethical aspects. First, we will seek ethical approval from the university ethical board.\footnote{We have already received preliminary approval for the design; we will seek approval for the final version resulting from this Registered Report.} Moreover, all activities related to surveys will be carried out entirely anonymously.
Participants will be informed that they have the right to withdraw from the experiment at any time.
Before starting the experiment, participants will be notified that all sessions will be recorded; they will be assured that this decision is not a reflection of any participant's performance, but rather serves to facilitate a more precise data analysis. After evaluation, all recordings will be deleted. 
Lastly, only gender data will be collected for statistical purposes and answering to this question will be completely optional.

\section{Threats to Validity}
This section gives an overview of the potential threats to validity in our controlled experiment~\cite{wohlin2012experimentation}.

Regarding \textit{internal validity}, it is crucial to ensure that students possess sufficient knowledge of the experimental environment. To mitigate this, we will recruit participants from our university by gathering data regarding their experience in software engineering. Additionally, we will inquire about their prior experience with the tools utilized in the experimental setting to eliminate any potential learning effect. 
However, participants could be biased when responding to the self-report questionnaire designed to measure learning experience. Better performance might occur in the treatment they prefer, or the subjective nature of the questionnaire could influence responses. To mitigate this, participants will be randomly assigned to both 2D and VR visualization tools without being informed about the hypothesis being tested. Furthermore, the questionnaire for evaluating the learning experience is based on a validated scale. Statistical tools and analyses will be employed, including assessments of distributions and summary metrics, to detect potential biases and outliers that might impact the results.
Before beginning the experiments, participants will be provided with a straightforward guide (in textual form) on using the setting.

Concerning \textit{external validity}, our experiment is inspired by the study conducted by Moreno-Lumbreras et al. \cite{moreno2024software}, which included 32 participants. To ensure comparable results, we plan to recruit at least the same number of participants, though a larger sample size, as discussed with the G*Power, would help to increase statistical power. Additionally, participant representation is a key consideration. Since we are working within an academic setting, we will use convenience sampling to recruit students who meet the necessary knowledge requirements for the experiment. To encourage proactive participation, each student will receive a personalized gadget as a token of appreciation. Additionally, 10 participants will be randomly selected to receive a €50 Amazon gift card.
Another potential challenge is ensuring similarity between the graphical interfaces of the two systems being tested. Since the study by Moreno-Lumbreras addressed this issue, we can maintain a similar level of interface consistency, thus supporting the validity of comparisons between the systems.

\balance


\end{document}